\documentclass{epl}
\usepackage{graphics}
\newcommand{\beq}{\begin{equation}}
\newcommand{\eeq}{\end{equation}}
\newcommand{\bdis}{\begin{displaymath}}
\newcommand{\edis}{\end{displaymath}}
\newcommand{\bea}{\begin{eqnarray}}
\newcommand{\eea}{\end{eqnarray}}
\newcommand{\barr}{\begin{array}}
\newcommand{\earr}{\end{array}}

\title{Short period attractors and non-ergodic behavior in the
deterministic fixed energy sandpile model}
 
\author{Franco Bagnoli\inst{1} \and Fabio Cecconi\inst{2} 
\and Alessandro Flammini\inst{3} \and Alessandro Vespignani\inst{4}}

\institute{
\inst{1} Dipartimento di Energetica "S. Stecco", 
         Via S. Marta 3, I-50139 Firenze, Italy \\
\inst{2} INFM and Physics Department, Universit\'a di Roma ``La
         Sapienza'', P.le A. Moro 2, I-00185  Rome, Italy \\
\inst{3} INFM and International School for Advanced 
         Studies (SISSA/ISAS), via Beirut 4, I-34014 Trieste, Italy \\
\inst{4} Laboratoire de Physique Th\'eorique (UMR du CNRS 8627),
         B\^atiment 210  Universit\'e de Paris-Sud 91405 ORSAY Cedex - France}
 
\pacs{05.70.Ln}{Nonequilibrium and irreversible thermodynamics} 
\pacs{05.65.+b}{Self-organized systems} 
\pacs{05.50.+q}{Lattice theory and statistics}

\begin{document}

\shortauthor{F.~Bagnoli \etal}
\shorttitle{Short period attractors in Fixed Energy Sandpiles}

\maketitle

\begin{abstract}
We study the asymptotic behaviour of the Bak, Tang, Wiesenfeld 
sandpile automata as a closed system with fixed energy. 
We explore the full range of energies characterizing the active phase. 
The model exhibits strong non-ergodic features by settling into 
limit-cycles whose period depends on the energy and initial conditions. 
The asymptotic activity $\rho_a$ (topplings density) shows, as a function 
of energy density $\zeta$, a devil's staircase behaviour defining a 
symmetric energy interval-set over which also the period lengths remain 
constant. 
The properties of $\zeta$-$\rho_a$ phase diagram can be traced back to 
the basic symmetries underlying the model's dynamics.
\end{abstract}

Sandpile automata and the associated self-organized 
critical scenario have been widely used to study the
occurrence of avalanche behavior in nature~\cite{jensen98}. 
These automata are, in general, driven-dissipative 
systems in which matter (sand) or energy is externally added to the
system and dissipated by the dynamics itself. 
Eventually, the input and output balance produces a stationary state with 
highly fluctuating bursts of activity, that in the limit of an infinitely 
slow input (time scale separation) is self-similar~\cite{grin,hwa,vz}. 
This condition is customarily embedded in the original 
sandpile model introduced by Bak, Tang and Wiesenfeld (BTW)
\cite{btw}, and the Manna model \cite{mannamodel}, which define
a deterministic and stochastic relaxation dynamics, respectively.

Recently, fixed energy sandpiles (FES) automata, 
which share the same microscopic dynamics of 
the corresponding BTW sandpiles without external driving and dissipation, 
have been studied \cite{fes,bigfes,cmv,mtak}. 
The energy density or sand is therefore a conserved quantity 
that acts as a tuning parameter.  
FES present an absorbing state phase transition (APT)~\cite{review}  
at a value of the energy density which is identical to the 
stationary energy density of the driven-dissipative 
version of the model\cite{bigfes}. FES with stochastic 
dynamics (Manna dynamics) belong to the universality class  
of  APT  with a conserved field\cite{rpv,lubeck} and 
the FES with deterministic BTW 
dynamics defines a different critical behavior with 
anomalies associated to non-ergodic effects\cite{bigfes}, 
similarly detected in the original driven model\cite{stella,kti}.

Interestingly, FES automata allows the study of the full 
phase diagram of sandpile automata. In particular, the 
overcritical phase is relevant with respect to experimental  
situations since many avalanche phenomena are naturally 
poised in this regime. Indeed, relations among sandpile automata, 
charge density waves and 
linear interface models have been established and prompt to
interesting behavior for the higher total energy properties
\cite{mid1,mid2,exp,alava}.

In this Letter we study the FES in order to 
provide a more detailed characterization of the model over the whole
energy range. The density of active sites shows a step-like  
behavior for increasing energies, with constant plateaus
in correspondence of energy intervals which form a hierarchical 
and symmetric interval set.
The model shows strong non-ergodic features, and after 
a transient relaxes onto periodic orbits which depend upon energy and
the initial conditions. Both the period lengths and  
transient times to reach the periodic ortbits remain constant onto the same 
energy intervals charactering the plateaus of the activity behavior. 
We tested the robustness of the observed behavior 
by looking at the scaling of activity and period lengths
with the system size.
As a preliminary understanding of some of the features observed 
in numerical simulations, we discuss analytically some basic symmetry
properties of the BTW toppling dynamics that account for the symmetry 
observed in the activity behavior and the corresponding energy 
intervals. 
The present results might also be relevant to  
intermittency and predictability issues in slowly driven 
BTW automata\cite{erzan,lpv} and could provide new insights
for the study of toppling invariants and recurrent states\cite{dhar}.

Here we consider the original BTW model in the square lattice
with periodic boundary conditions. 
The configuration is specified by giving the {\it energy},
$z_i$, at each site.  The energy may take integer values, 
and is nonnegative in all cases.
Each active site, i.e.,  with (integer) energy greater than or 
equal to the {\em activity threshold} $z_{th}$ ($z_i \geq z_{th} = 4$), 
topples and redistributes its energy following the updating rules
 $z_i \rightarrow z_i - z_{th}$, and $z_j \rightarrow z_j + 1$ at
each of the $4$ nearest neighbors of $i$. 
The BTW dynamics with {\it parallel} updating
(all active sites topple at each update) is completely deterministic
and Abelian; i.e. the order
in which active sites are updated is irrelevant
in the generation of the final (inactive) configuration \cite{dhar}.

In the FES, the energy density $\zeta=L^{-2}\sum_i z_i$ is a conserved 
quantity since no external driving 
is present and periodic conditions are assumed at the boundaries of the 
lattice (no dissipation). 
The value of $\zeta$ is fixed by the initial condition 
which is generated by distributing $\zeta L^2$ particles randomly among 
the $L^2$ sites.
Once particles have been placed, the dynamics begins.
If after some time the system falls into a configuration with 
no active sites, the dynamics is permanently frozen, i.e., the 
system has reached an absorbing configuration. By varying $\zeta$, 
FES show a phase transition separating an absorbing phase (in which  
the system always encounters an absorbing configuration), 
from an active phase with sustained activity. 
This has been assumed to be a continuous 
phase transition, at which the system shows critical behavior.  
The order parameter is the stationary average density of 
active sites $\rho_a$, which equals zero for $\zeta\le \zeta_c$.
Previous studies focused on the behavior close to the critical
point in order to characterize the scaling behavior 
$\rho_a\sim(\zeta-\zeta_c)^\beta$, for $\zeta>\zeta_c$ \cite{bigfes}.
Numerical results, however, pointed out possible failures of 
simple scaling hypothesis and nonergodic behavior as also observed in
slowly driven simulations\cite{stella,kti}. These features contradict to
the standard scaling observed in stochastic sandpile
models \cite{bigfes}. 

\begin{figure}
\begin{center}
\includegraphics[clip=true,width=0.65\columnwidth, keepaspectratio]
{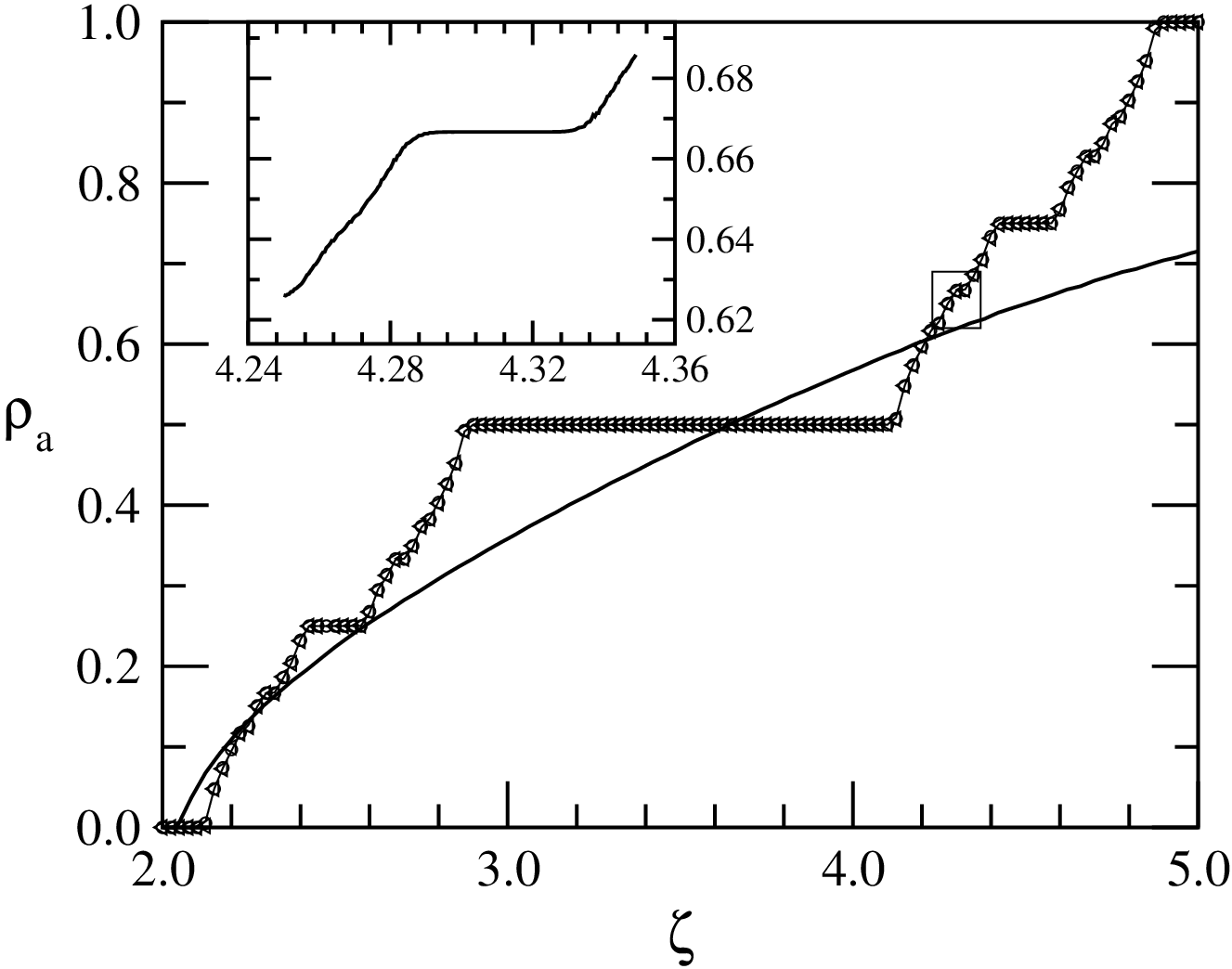}
\end{center}
\caption{Stationary density $\rho_a=n_T/T$ vs. energy density for
lattice sizes: 
$L=50$ (square), $L=100$ (open circles), $L=200$ (triangles).
Data are averaged over $N=200$ random initial conditions.
The solid line curve represents the activity $\rho_a$ for the
stochastic Manna model ($z_{th}=4$) with size $L=100$.
Inset: enlargement of the region $4.23<\zeta<4.26$ (box) showing that a
smaller plateaus are observed on finer scales.}
\label{fig1}
\end{figure}

In order to exploit the effect of the deterministic BTW dynamics 
we consider the active sites density behavior for the whole range of  
$\zeta>\zeta_c$; i.e. the active phase.
Since for any finite size, the number of possible configurations
is finite and the dynamics is deterministic, after a  
transient the system enters a periodic orbit, visiting recursively a 
finite set of configurations. In such a case, $\rho_a$ can 
be computed as the ratio of the total number of topplings at a site 
in a period and the length of the period itself.
In Fig.\ref{fig1} we plot the activity density $\rho_a$ in the
stationary state as a function of the energy $\zeta$.  

The activity has a stairway structure with constant activity 
plateaus over energy intervals distributed symmetrically with 
respect to the energy value $\zeta=3.5$. The step-like 
behavior is in contrast with the smooth and regular curve obtained 
in the case of a stochastic models in which, for each toppling event,  
energy grains are redistributed at random between
neighbors\cite{mannamodel}.
The value of $\rho_a$ corresponding to each value of $\zeta$ is the
average over at least $N=200$ initial conditions (IC).
In the activity plateaus we do not observe a dependence of the stationary
activity with respect to IC. For energies in the central 
and larger plateau, for instance, all initial conditions
evolve to an orbit with period 2 and a single toppling per site
per period for large lattice size. Correspondingly, the
observed value of activity is  delta-like distributed at
$\rho_a=1/2$. 
On the contrary, for energies outside of the plateaus, we find that 
different IC's may generate different values 
of the stationary $\rho_a$ which exhibits a moderate dispersion around 
the mean reported in the plot of Fig.~\ref{fig1}. 
The mean value is not recovered in individual runs, signalling the 
non-ergodic properties already noticed in
Ref.~\cite{bigfes}. It is worth remarking that this scenario
presents striking analogies with complex 
phase-locking plateaus characterized by devil's staircases 
found in numerous non-linear driven systems\cite{pl}.

\begin{figure}
\begin{center}
\includegraphics[clip=true,width=0.65\columnwidth, keepaspectratio]
{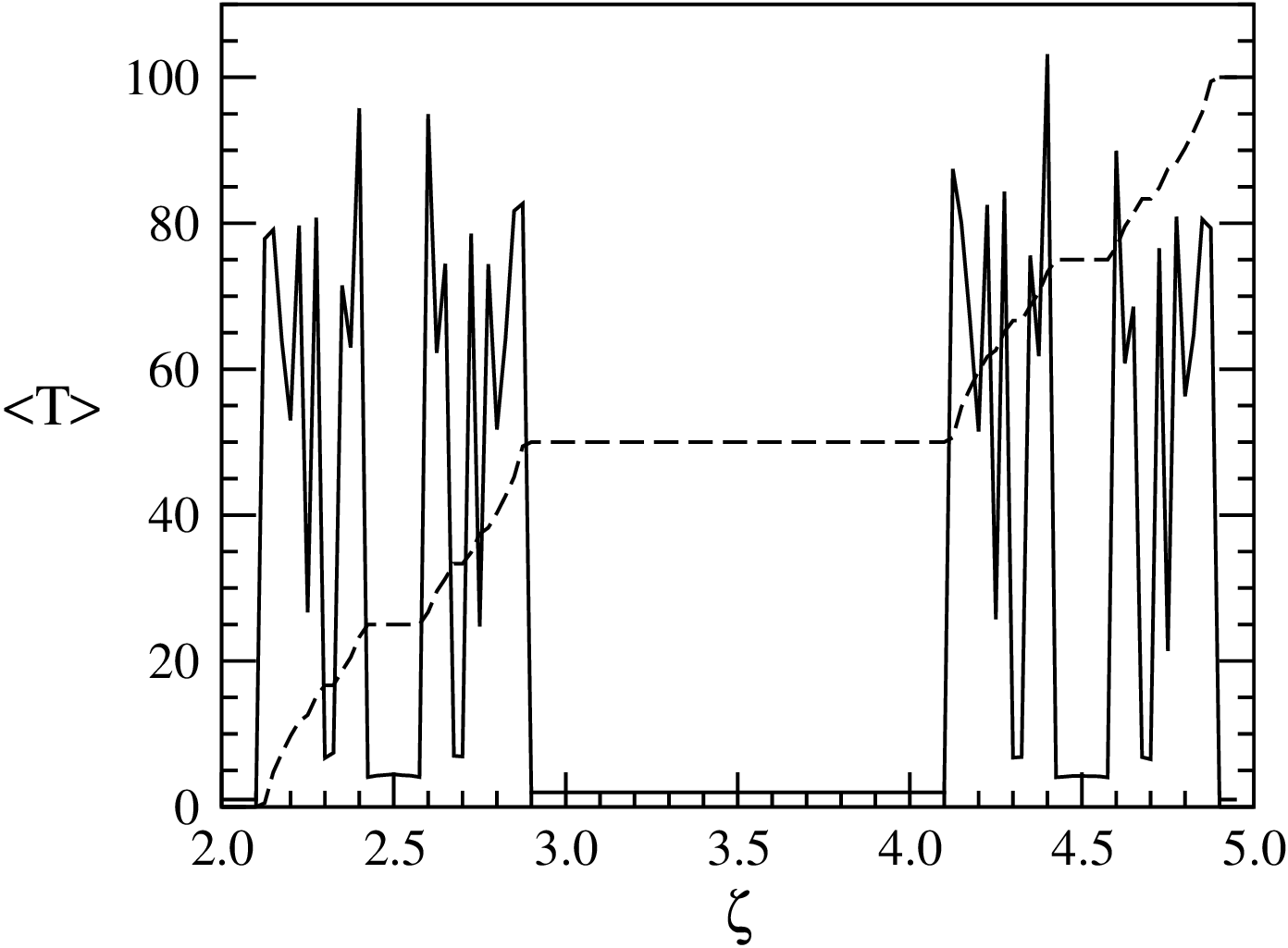}
\end{center}
\caption{Behavior of average period $\langle T \rangle$ as a function of
energy density for system size $L=100$.
The average is performed over at least $10^3$ randomly chosen initial
conditions.
Dashed line indicates the corresponding stationary $\rho_a$ after a
suitable rescaling for a direct comparison.}
\label{fig2}
\end{figure}

A confirmation of the non-ergodic nature of the BTW model 
is  provided by measuring, as a function of energy, the length 
of the periodic cycle where the same sequence of topplings is 
executed over and over (periodic orbit). 
The period lengths depend on the initial condition and 
on the system's energy. 
In Fig.~\ref{fig2} we plot the average period length $\langle T \rangle$ as a 
function of $\zeta$ for a system of size $L=100$.  
In the same figure $\rho_a$ (suitably rescaled) is shown 
to better appreciate the very same  correspondence of the 
plateau structures. On the plateaus the cycles period 
assumes extremely low values, such as $T=2,4,6$. As for the activity
$\rho_a$, the period lengths are delta distributed in correspondence 
of the plateaus; i.e. all IC define the same $T$. 
Outside the energy plateaus, different IC generate different orbits
defining a nontrivial  period distribution.   
It is interesting to note that a similar irregular behavior of periods 
is found in other deterministic dissipative systems with continuous state 
variables which exhibit unpredictable transient dynamics but admit asymptotic 
periodic states as attractors \cite{Cecco}. 
The robustness of the present picture has been tested through simulations
of systems of increasing size. In Fig.\ref{fig1} we report  data 
for system sizes $L=50, 100$ and $200$. The discrete nature of the system 
allows to discriminate energies on a finer scale ($1/L^2$), and the 
structure of the energy intervals defined by activity plateaus reveals 
progressively a reacher structure \cite{note1}.
With data at hand we cannot say whether, in the infinite size limit, 
the system is characterized by an energy range
made exclusively by infinitesimal intervals and correspondingly,
the curve $\rho_a(\zeta)$ is discontinuous.
In this case also the transition inactive-active would 
be discontinuous, ruling out the possibility even to define 
the critical exponent $\beta$ characterizing the transition.

\begin{figure}
\begin{center}
\includegraphics[clip=true,width=0.65\columnwidth, keepaspectratio]
{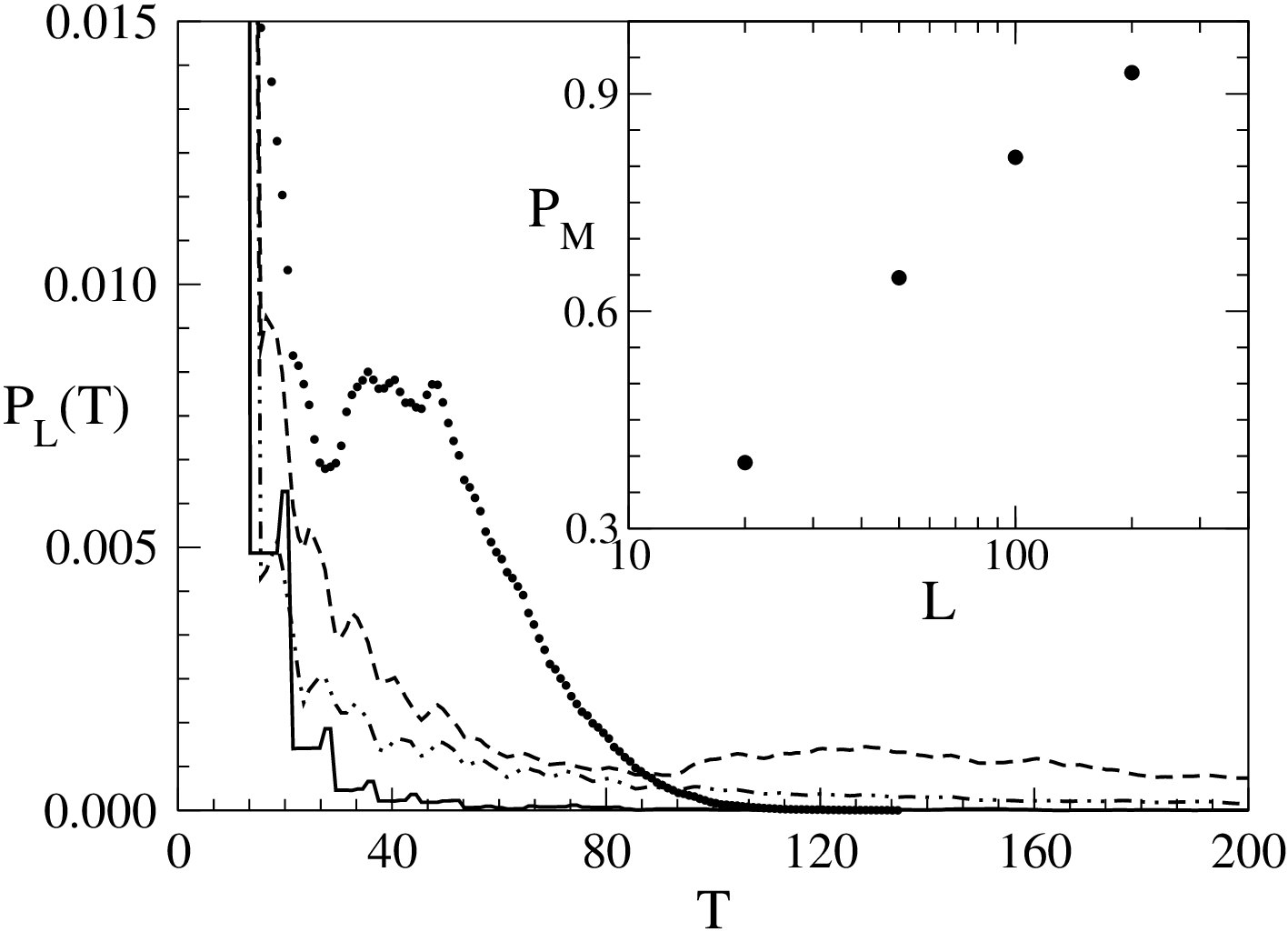}
\end{center}
\caption{Histograms of periods collected in the stationary states 
at the energy density $\zeta = 4.25$ and different system sizes:
$L=20$ (circles),$L=50$ (dashed),$L=100$ (dot-dashed) and $L=200$ (solid).
The maxima $P_m$ of the histogram, corresponding to $T=8$, lie outside of the 
graph.  The inset shows the increasing of $P_m$ with the system size
$L$, this numerical evidence seems to support the view that 
period selection becomes stronger as the size increases.}   
\label{fig3}
\end{figure}

It is also worth discussing the non-ergodic effect with respect to IC.
This feature does not disappear for increasing sizes, even though a 
single specific value for the period and the number of topplings 
per period (and therefore for $\rho_a$) emerges with overwhelming frequency.   
As an illustration of the basic phenomenology, we report in
Fig.~\ref{fig3} the normalized histograms of periods $T$ collected from
a sample of $10^3$ random IC, at different sizes and $\zeta=4.25$. 
The results indicate that, as the size $L$ grows, each distribution seems to 
becomes more and more peaked around a given period value $T_M$ 
($T_M=8$ for $\zeta=4.25$) and the inset of Fig.~\ref{fig3} 
shows the increase of the histogram maximum $P_M$ (peak) with $L$.    
This could be consistent with a period selection scenario in which, in
the thermodynamic limit, all IC select a single period length. 
Finite size effects, however, are still too large to 
conclude definitely that such a mechanism occurs for $L\to\infty$. 

A first understanding of the behavior of the fixed energy BTW model 
can be achieved analytically by exploiting the symmetry 
contained in the deterministic dynamical rules. The 
non-ergodicity with respect to the IC can be traced back to 
conserved invariants of the dynamics. For instance, 
the initial configuration in which all grains are concentrated in 
the central site of the lattice will necessarily evolve preserving 
its central symmetry and asymmetric configurations will never be visited. 
Other quantities conserved by the dynamics can be constructed
mimicking the center of mass and the angular momentum of an isolated 
Hamiltonian system. 
A straightforward calculation shows that
\begin{equation}
\begin{array}{c}
MX = \sum_{x,y} x z_{x,y}  \\
                           \\
MY = \sum_{x,y} y z_{x,y}  \\
                            \\
MXY = \sum_{x,y} (x^2 - y^2) z_{x,y}
\end{array}
\end{equation}
are dynamical invariants if taken modulus the linear size of the lattice,
with $x$ and $y$ indicating the integer coordinates of the lattice sites with
respect to the origin ($x=0,y=0$).
The phase space relative 
to a given value of the energy is therefore partitioned according 
to the values assumed by conserved quantities~\cite{note2}. 
A detailed and comprehensive discussion on toppling invariants can be 
found in Ref.~\cite{DharCM}. 
It is worth noticing, that the previous considerations hold for
any finite system but it is not possible to generalize them to the
thermodynamic limit. 

A striking symmetry property of the model concerns the distribution
of toppling per site within  any given periodic orbit.
In order to exploit this property let $n_i$ be the number of times that
the site $i$ topples during a period. Since after a period the initial
configuration is restored, each site has given to its neighbors as many 
grains as it has received, which translates in:

\begin{equation}
4 n_{i}=\sum_{j \in nn} n_j,
\end{equation}
\noindent
where $j \in nn$ denote the sum over the nearest neighbors of the
site $i$. Let us suppose that not all $n_{i}$ are the same 
and define $i^*$ the site with the minimum number of topplings
in a period $n_{i^{*}}$. 
The quantity  $p_{i}=n_{i}-n_{i^{*}}$ satisfies the same equation as 
the $n$'s and $p_{i^{*}}=0$ is the minimum $p$. 
Since $p_{i^{*}}$ is the sum of four $p$'s, that, 
by definition are nonnegative we have that all $p$'s in the 
neighbors of $i^{*}$ are zero. Iterating the argument, the 
entire lattice can be covered, and one obtains that all $p$'s are zero and 
therefore all $n$'s are the same. This results immediately implies
that any site topples exactly the same number $n_T$ of times over 
a given period of length $T$ and eventually leads to the interesting 
result 
\begin{equation} 
\rho_a  = \frac{n_T}{T}. 
\end{equation}
Finally, we can focus on the evident symmetries of the 
activity behavior $\rho_a (\zeta)$ and the corresponding energy
intervals. The overall symmetric structure of the energy intervals 
centered at $\zeta=3.5$ can be accounted for by the following argument.
Let us consider two energy fields $z_i(t)$ and $z'_i(t)$,
that at time $t$ are related as 
$z_i(t)=7-z'_i(t)$, but otherwise arbitrary.
It can be easily verified that the above relation holds at all 
subsequent times. 
In fact, the evolution equation for $z_i$'s can be written as 
\begin{equation}
z_{i}(t+1)= z_{i}(t)
+ \sum_{j \in nn}\Theta[z_j(t) \ge 4 ] - 
4\Theta[z_i(t) \ge 4 ],
\end{equation}
and by substituting $z_i(t)\to 7-z'_i(t)$ we obtain 
\begin{equation}
z_i(t+1)= 7 - z'_i(t)
+\sum_{j \in nn}\Theta[z'_{j}(t) \le 3 ] - 
4\Theta[z'_i(t) \le 3 ].
\end{equation}
By using the relation $\Theta[g \le 3] =1-\Theta[g \ge 4]$ 
we readily obtain 
\begin{equation}
z_{i}(t+1)= 7 - z'_{i}(t+1),
\end{equation}
This show recursively that if at a given time $t$ the two fields 
are related by $z_i(t)=7-z'_i(t)$ they are similarly related at 
all subsequent times. In addition if a site topples in one of 
the two lattices, the corresponding site in the other won't, 
and vice versa leading to $\rho_a(z_i)= 1-\rho_a(z'_i)$.
The full account of the observed symmetry under transformation 
$\zeta \rightarrow 7-\zeta$ and $\rho_a(\zeta) \rightarrow 
1-\rho_a(7-\zeta)$ would require that an equal proportion of IC
settle into configurations related by $z'_i=7-z_i$.
While it is not possible to show analytically the latter statement, 
the symmetry is recovered exactly in the diagram of Fig.~\ref{fig1}.

Despite these simple symmetry arguments account for several features
of the BTW model, many other issues remain unsettled. In particular,
the activity and period length plateaus extending over the energy
intervals do not find a simple analytical explanations. As well,
the continuous or discontinuous nature of the $\rho_a(\zeta)$ behavior 
cannot be discriminated by the simple analytical arguments provided
here.

In summary, we presented a numerical and analytical study of the 
BTW automaton with fixed energy. We find strong non-ergodic features
related to the deterministic dynamics of the model in the whole energy
range. Some features of the model can be understood in terms of the
basic symmetries of the dynamics. A full rationalization of the present 
results could help to understand the scaling anomalies and the 
universality class of deterministic  self-organized critical  
sandpile models.


\begin{thebibliography}{0}
\bibitem{note1}
	In this sense, the energy interval set 
	seems to have scaling properties typical of Cantor-like
	fractal sets. A precise characterization this structure is
	beyond the scope of the paper and will be addressed in future works.
\bibitem{note2}
	Note that the conservation laws discussed above rely only 
	upon the simmetry of the toppling events and not on the 
	fact that the dynamics is threshold-activated.
\end{thebibliography}
\end{document}